\documentclass[11pt]{article}

\usepackage[margin=.75in]{geometry}
\usepackage{datetime}
\usepackage{amsmath, amsthm, amssymb,fancyhdr,mathrsfs,  enumerate}
\usepackage{latexsym}
\usepackage[round, authoryear, comma, sort&compress]{natbib}
\usepackage{hyperref}
\usepackage[utf8]{inputenc}
\usepackage{graphicx,float}
\usepackage[export]{adjustbox}

\newtheorem{lemma}{Lemma}

\renewcommand{\P}{\mathbb{P}}

\newcommand{\bpsi}{{\bf\Psi}}
\newcommand{\bb}{{\bf\beta}}

\newcommand{\E}{\mathbb{E}}
\newcommand{\I}{\mathbb{I}}
\newcommand{\R}{\mathbb{R}}

\newcommand{\widesim}[2][1.5]{
  \mathrel{\overset{#2}{\scalebox{#1}[1]{$\sim$}}}
}

\begin{document}

\begin{center}
      \Large{\bf {Liu-type Shrinkage Estimators for Mixture of Logistic Regressions: An Osteoporosis Study}}
\end{center}
\begin{center}
 \noindent{{\sc Elsayed Ghanem$^{\dagger,\ddagger}$},  
 {\sc Armin Hatefi$^{\dagger,}$\footnote{Corresponding author:\\
 Email: ahatefi@mun.ca and Tel: +1 (709) 864-8416}}
  and {\sc Hamid Usefi$^{\dagger}$}}

\vspace{0.5cm}
\noindent{\footnotesize{\it $^{\dagger}$Department of Mathematics and Statistics, Memorial University of Newfoundland, St. John's, NL, Canada.}} \\
\noindent{\footnotesize{\it $^{\ddagger}$Faculty of Science, Alexandria University, Arab Republic of Egypt.}} \\
\end{center}

\begin{center} 
{\small \bf Abstract} 
\end{center}

The logistic regression model is one of the most powerful statistical methods
 for analysis of binary data. The logistic regression
 allows to use a set of covariates to explain the binary responses. 
 The mixture of logistic regression models is used to fit heterogeneous populations through an unsupervised learning approach. 
 The multicollinearity problem is one of the most common problems in logistic and mixture of logistic regressions where the 
 covariates are highly correlated. This problem results in unreliable maximum likelihood estimates for the regression coefficients. 
This research developed shrinkage methods to deal with the multicollinearity in a mixture of logistic regression models. 
These shrinkage methods include ridge and Liu-type estimators. Through extensive numerical studies, we show that the developed
 methods provide more reliable results in estimating the coefficients of the mixture. Finally, we applied the shrinkage methods to analyze the bone disorder status of women aged 50 and older.   


\noindent {\bf Keywords: } Multicollinearity, Maximum likelihood, Ridge penalty, Liu-type penalty, Logistic regression, 
Mixture models, EM algorithm, Bone mineral data.

\section{ Introduction } \label{sec:intro}
Osteoporosis is a bone disorder that occurs when the bone architecture of the body dramatically declines.
This deterioration leads to various major health issues. Patients with osteoporosis, for example, 
are more susceptible to skeletal fragility and fractures e.g., in the spine, hip and femur areas \citep{cummings1995risk,melton1998bone}. 
Osteoporosis has a substantial impact on a patient's health and survival. More than half of patients 
suffering from osteoporotic hip fractures will not be able to live independently and approximately
one-third of these patients will die within one year from the medical complication of the disease \citep{bliuc2009mortality,neuburger2015impact}.
 The financial burden of osteoporosis is also undeniable on community health. \cite{lim2016comparison}, for 
 instance, reports that the annual cost of osteoporosis and its related health problems is  
 twice as much as that of diabetes in South Korea.  

According to the WHO expert panel, bone mineral density (BMD) is considered the most reliable factor 
in diagnosing osteoporosis \citep{world1994assessment}. The bone status is determined as osteoporosis when the BMD scores 
less than 2.5 SDs from the BMD norm (i.e., the BMD mean of healthy individuals between 20 and 29).
 The density of bone tissues increases until age group 20-30 and then decreases as the individual gets older.
  In addition to age, there are various research articles in the literature studied the association between osteoporosis 
  and characteristics of patients, such as sex, weight, and BMI \citep{lloyd2014body,kim2012relationship,de2005body}.       

The logistic regression model is one of the most popular statistical methods to model the association between 
covariates  (e.g., patients' characteristics) and binary responses (e.g., osteoporosis status of the patient). 
Maximum likelihood (ML) is a standard method for estimating the parameters of the logistic regression model. The ML 
method requires no restriction on the set of covariates in estimating the logistic regression coefficients. 
Despite the flexibility, the ML estimates are significantly affected by multicollinearity, where the covariates 
are linearly dependent.  \cite{schaefer1984ridge} incorporated the ridge penalty in logistic regression estimation and proposed ridge 
logistic regression to cope with the multicollinearity issue. 
When a multicollinearity problem is severe, the ridge estimator may not be able to address the ill-conditioned design matrix. 
\cite{liu2003using} proposed the Liu-type (LT) penalty for the linear regression model to control the bias of the ridge method and 
handle the high multicollinearity. \cite{inan2013liu} investigated the ridge and LT methods in estimating the 
coefficients of the logistic regression model. \cite{pearce2021multiple} recently developed ridge and 
LT shrinkage estimators under ranked set sampling designs 
for logistic and stochastic restricted regression models. 

Finite mixture models (FMMs) provide a powerful and convenient tool to model mathematically populations 
consisting of several subpopulations. \cite{quandt1978estimating} extended the idea of FMMs to linear regression 
models and introduced a mixture of linear regression models. Expectation-maximization (EM) algorithm \citep{dempster1977maximum} 
is a well-known technique to find the ML estimates of FMMs and a mixture of logistic regression models. 
\cite{aitkin1999general,aitkin1999meta,wang1998mixed} used the ML method to estimate the parameters of a finite mixture of logistic regression models.  
\cite{celeux1992classification,celeux1985sem} developed stochastic versions of the EM algorithm to find the ML estimates of the mixture parameters. 
Mixture models have found applications in the core of statistical sciences, such as classification and modelling data 
from various sampling structures, including stratified sampling \citep{wedel1998mixture} and ranked set sampling \citep{hatefi2015mixture,hatefi2018} to 
name a few. Readers are referred to \citep{peel2000finite} for more details about the theory and applications of the FMMs.

In this paper, we focus on the finite mixture of logistic regression models. Similar to logistic regression,
 the ML estimates of the mixture of
logistic regressions are severely affected by multicollinearity. We developed the LT shrinkage 
estimator for the mixture of logistic regression models. Through 
 various simulation studies, we show  the LT estimators outperform their
 ridge and ML counterparts in estimating the coefficients of the mixture of logistic regressions. The estimation methods are finally applied to bone mineral data to analyze 
 the bone disorder status of women aged 50 and older.  

This paper is organized as follows. Section \ref{sec:meth} describes the ML, ridge and LT methods in estimating the parameters 
of the mixture of logistic regression models. 
Sections \ref{sec:sim} and \ref{sec:real} assess the performance of the estimation methods via various simulation studies and a 
real data example.  The summary and concluding remarks are finally presented in Section \ref{sec:sum}.

\section{Statistical Methods}\label{sec:meth}

Logistic regression is considered as one of the most common statistical tools for analysis 
of binary responses. 
Let ${\bf y}=(y_1,\ldots,y_n)$ denote the vector of binary responses from a sample of size $n$.
Let ${\bf X}$ denote $(n \times p)$ design matrix of $p$ explanatory variables $({\bf x}_1,\ldots,{\bf x_p})$
of $\text{rank}({\bf X})=p < n$.
Given $({\bf X},{\bf y})$, the logistic regression model is given by
\begin{align}\label{logistic}
\P(y_i=1|{\bf X}) = g^{-1}({\bf x}_i;{\bf \beta})= 
1/ \left(1 + \exp(-{\bf x}^\top_i {\bf \beta})\right),
\end{align}
 where $g$ denotes the link function and ${\bf \beta}$ represents the vector 
of unknown coefficients. The logistic regression \eqref{logistic} aims to model the association between 
explanatory variables $({\bf x}_1,\ldots,{\bf x_p})$ with  the binary response $y_i \in \{0,1\}$ 
observed from $i$-th subject for $i=1,\ldots,n$. 

The ML method is the most common approach to estimate the coefficients 
of the logistic regression. 
To obtain the ML estimate of ${\bf \beta}$, we first require the likelihood function of the coefficients
 given the observed data.  From the Bernoulli distribution of the responses, the 
 log-likelihood function of ${\bf \beta}$ is given by
 \begin{align}\label{loglik_logistic}
 \ell({\bf \beta}) = \sum_{i=1}^{n} 
 \left\{
 y_i {\bf x}^\top_i {\bf \beta}  - \log \left(1 + \exp(-{\bf x}^\top_i {\bf \beta})
\right)
 \right\}.
 \end{align}
 
As a generalization of logistic regression \eqref{logistic},  the mixture of logistic regression 
models is used when 
the population of interest comprises several subpopulations (henceforth called components). 
Let $M$ denote the umber of the components of the mixture of logistic regression models. 
While we assume that the number of components is known in this mansucript, the problem of mixture of logistic regression is 
treated as an unsupervised learning approach where the component membership of the observations are unknown 
and must be estimated. From \eqref{loglik_logistic}, the log-likelihood of the mixture of logistic 
regressions  follows
 \begin{align}\label{loglik_mix}
 \ell(\bpsi) = \sum_{i=1}^{n} \log
 \left\{ \sum_{j=1}^{M} \pi_j [p_j({\bf x}_i;\bb_j)]^{y_i} [1-p_j({\bf x}_i;\bb_j)]^{(1-y_i)}
 \right\},
 \end{align}  
where 
\begin{align}\label{pj}
 p_j({\bf x}_j;\bb_j) = g^{-1}({\bf x}_i;\bb_j),
  \end{align}
  and ${\bf \pi}=(\pi_1,\ldots,\pi_M)$ represents the vector of the mixing proportions with $\pi_j >0$ and 
  $\sum_{j=1}^{M} \pi_j=1$. Also,
  we use $\bpsi=({\bf \pi},{\bb})$ with ${\bb}=(\bb_1,\ldots,\bb_M)$ to represent the vector of
  all unknown parameters of the mixture.  
  
\subsection{ML Estimation Method}\label{sub:ls}

  There is no closed form for the maximizer of the log-likelihood function  \eqref{loglik_mix} in 
  estimating the parameters of the mixture model. Thus, we view  $\{({\bf x}_i, {y}_i), i=1,\ldots,n\}$ 
  as an incomplete data and develop an expectation-maximization (EM) algorithm \citep{dempster1977maximum} to obtain the ML estimate of $\bpsi$. 
 Suppose $\{({\bf x}_i, {y}_i,{\bf Z}_i), i=1,\ldots,n\}$ denote the complete data where ${\bf Z}_i = (Z_{i1},\ldots,Z_{iM})$ 
 is a latent variable representing the component membership of the $i$-th subject with
  \begin{align*}
 Z_{ij} = \left\{
 \begin{array}{lc}
 1 & \text{if the $i$-th subject comes from the $j$-th component,}\\
 0 & o.w.
 \end{array} \right.
 \end{align*}  
Given ${\bf Z}_i \widesim{iid} \text{Multi}(1,\pi_1,\ldots,\pi_M)$, the joint distribution of $(y_i,{\bf Z}_i)$ can be written as
\begin{align}\label{f_y.z}
 f(y_i,{\bf Z}_i) = 
 \prod_{j=1}^{M} 
 \left\{ \pi_j
 [p_j({\bf x}_i;\bb_j)]^{y_i} [1-p_j({\bf x}_i;\bb_j)]^{(1-y_i)} \right\}^{z_{ij}}.
 \end{align}  
 From above, it is easy to show   ${\bf Z}_i|y_i \widesim{iid} \text{Multi}(1, \tau_{i1}(\bpsi),\ldots,\tau_{iM}(\bpsi))$ where
 \begin{align}\label{tau_ls}
 \tau_{ij}(\bpsi) = \frac{\pi_j [p_j({\bf x}_i;\bb_j)]^{y_i} [1-p_j({\bf x}_i;\bb_j)]^{(1-y_i)}}{ 
 \sum_{j=1}^{M}\pi_j [p_j({\bf x}_i;\bb_j)]^{y_i} [1-p_j({\bf x}_i;\bb_j)]^{(1-y_i)}}.
 \end{align} 
 Using the latent variables ${\bf Z}_i$, the complete log-likelihood function of $\bpsi$ is given by 
 \begin{align}\label{log_like_comp}
 \ell_c(\bb) = \sum_{i=1}^{n} \sum_{j=1}^{M} z_{ij} \log(\pi_j)
 + \sum_{i=1}^{n} \sum_{j=1}^{M} z_{ij} \log \left\{ [p_j({\bf x}_i;\bb_j)]^{y_i} [1-p_j({\bf x}_i;\bb_j)]^{(1-y_i)} \right\}.
\end{align}

 EM algorithm decomposes the estimation procedure into two iterative steps, including expectation (E-step) and
 maximization (M-step). In this manuscript, we use stochastic EM (SEM) algorithm \citep{celeux1985sem} to estimate the parameters 
 of the mixture of logistic regressions. 
 The SEM algorithm re-designs the EM algorithm and accommodates a stochastic classification step (S-step) between E- and M-steps. 
 
 As an iterative  method, SEM algorithm requires starting points to initiate the estimation process. 
 Let $\bpsi^{(0)} = ({\bf\pi}^{(0)},\bb^{(0)})$ denote the starting points. 
 In the following, we describe how E-, S- and M- steps  are implemented in the $(l+1)$-th 
 iteration when $\bpsi^{(l)}$ represents the the update from $l$-th iteration. 
 
{\bf E-Step}: One first requires to compute the conditional expectational of latent variables given incomplete data. Hence,  
\[
\E_{\bpsi^{(l)}} (Z_{ij}|y_i) = \tau_{ij}(\bpsi) |_{\bpsi=\bpsi^{(l)}}= \tau_{ij}(\bpsi^{(l)}),
\]
where  $\tau_{ij}(\bpsi^{(l)})$ is calculated from \eqref{tau_ls}.
Then, the conditional
expectation of the log-likelihood function \eqref{log_like_comp} can be re-written by 
\[
{\bf Q}(\bpsi,\bpsi^{(l)}) = \E_\bpsi (\ell_c(\bb) | {\bf y}, \bpsi^{(l)}) = {\bf Q}_1({\bf \pi},\bpsi^{(l)})  + {\bf Q}_2({\bb},\bpsi^{(l)}), 
\] 
where 
\begin{align} \label{Q1_ls}
{\bf Q}_1({\bf \pi},\bpsi^{(l)}) = \sum_{i=1}^{n} \sum_{j=1}^{M} \tau_{ij}(\bpsi^{(l)}) \log(\pi_j),
\end{align}
and 
\begin{align} \label{Q2-ls}
{\bf Q}_2({\bb},\bpsi^{(l)}) = \sum_{i=1}^{n} \sum_{j=1}^{M} \tau_{ij}(\bpsi^{(l)}) 
\log \left\{ [p_j({\bf x}_i;\bb_j)]^{y_i} [1-p_j({\bf x}_i;\bb_j)]^{(1-y_i)} \right\}.
\end{align}

{\bf S-Step}: We partition the subjects into  ${\bf P}^{(l+1)}=(P_1^{(l+1)},\ldots,P_M^{(l+1)})$ 
based on a stochastic assignment $( {Z}^*_{i1},\ldots,{Z}^*_{iM})$, given their posterior probability 
memberships $(\tau_{i1}(\bpsi^{(l)}),\ldots,\tau_{iM}(\bpsi^{(l)}))$. In other words, we generate 
${\bf Z}^*_i \widesim{iid} \text{Multi}(1,\tau_{i1}(\bpsi^{(l)}),\ldots,\tau_{iM}(\bpsi^{(l)})$
and the $i$-th subject is then classified to $P_h^{(l+1)}$ when ${Z}^*_{ih}=1$
for $i=1,\ldots,n$. 
In addition, if one of the partitions becomes empty or ends up with only one subject,  
the SEM algorithm is stoped and $\bpsi^{(l)}$ is retuned. 

{\bf M-Step}: In this step, we use the ${\bf P}^{(l+1)}$ of the S-step to update $\bpsi$. First, we maximize ${\bf Q}_1({\bf \pi},\bpsi^{(l)})$ from \eqref{Q1_ls} subject to constraint 
$\sum_{j=1}^{M} \pi_j=1$. Using the Lagrangian multiplier, it is easy to see 
\begin{align}\label{Jpihat_ls}
{\widehat \pi}_j^{(l+1)} = \sum_{i=1}^{n}  z^*_{ij} /n = n_j/n;  ~~~~ j=1,\ldots,M-1,
\end{align}
where $n_j$ denotes the number of subjects classified to ${P_j}^{(l+1)}$.
To estimate the coefficients of the $j$-th logistic regression, one can re-write \eqref{Q2-ls} based 
on partition ${\bf P}^{(l+1)}$ as follows
\begin{align}\label{JQ2_ls}
{\bf Q}_2(\bb_j,\bpsi^{(l)}) = \sum_{i=1}^{n_j} \tau_{ij}(\bpsi^{(l)}) 
 \left(y_i {\bf x}_i^\top \bb_j - \log \left( 1+ \exp(-{\bf x}_i^\top \bb_j) \right)
 \right).
\end{align}
From the first derivative of \eqref{JQ2_ls}, the gradient is given by
\begin{align}\label{grad_ls}
\nabla_{\bb_j} {\bf Q}_2(\bb_j,\bpsi^{(l)}) = {\bf X}_j^\top \left( {\bf y}_j - {\bf g}^{-1} ({\bf X}_j;\bb_j^{(l)}) \right)
\end{align}
where ${\bf X}_j$ and ${\bf y}_j$ are respectively the design matrix and vector of responses 
corresponding to subjects from ${P_j}^{(l+1)}$. Also, 
${\bf g}^{-1} ({\bf X}_j;\bb_j^{(l)}) = \left( {g}^{-1} ({\bf x}_1;\bb_j^{(l)}),\ldots, {g}^{-1} ({\bf x}_{n_j};\bb_j^{(l)})\right)^\top$
 where $g^{-1}(\cdot,\cdot)$ is given by \eqref{logistic}. The Hessian matrix of \eqref{JQ2_ls} is then obtained as
 \begin{align}\label{hess_ls}
{\bf H}_{\bb_j}\left({\bf Q}_2(\bb_j,\bpsi^{(l)})\right) = - {\bf X}_j^\top {\bf W}_j {\bf X}_j,
\end{align}
where  ${\bf W}_j$ is a diagonal matrix with entries 
 \begin{align}\label{w_ls}
(w)_{ii} = \exp({\bf x}_i^\top \bb_j^{(l)}) \left[1+\exp({\bf x}_i^\top \bb_j^{(l)})\right]^{-2}.
\end{align}
From \eqref{grad_ls} and \eqref{hess_ls}, one can use Newton-Raphson (NR) method and update $\bb_j, j=1,\ldots, M$ as follows 
 \begin{align}\label{nr_ls}
\bb_j^{(l+1)} = \bb_j^{(l)} - {\bf H}_{\bb_j}^{-1}\left({\bf Q}_2(\bb_j,\bpsi^{(l)})\right) {\nabla_{\bb_j} {\bf Q}_2(\bb_j,\bpsi^{(l)})}.
\end{align}

\begin{lemma}\label{le:ls}
Let $\nabla_{\bb_j} {\bf Q}_2(\bb_j,\bpsi^{(l)})$ and ${\bf H}_{\bb_j}\left({\bf Q}_2(\bb_j,\bpsi^{(l)})\right)$ 
 denote the gradient and Hessian matrix of \eqref{JQ2_ls}. Then the iteratively re-weighted least square 
  (IRWLS) estimate of $\bb_j$ can be obtained by
\[
{\widehat \bb}_j = \left({\bf X}_j^\top {\bf W}_j {\bf X}_j\right)^{-1} {\bf X}_j^\top {\bf W}_j {\bf V}_j,
\]
where ${\bf W}_j$ is diagonal weight matrix from \eqref{w_ls} and 
\[
{\bf V}_j = \left\{  {\bf X}_j {\widehat \bb}_j^{(l)}  +  {\bf W}_j^{-1} \left[ {\bf y}_j -  {\bf g}^{-1} ({\bf X}_j;{\widehat \bb}_j^{(l)})\right] \right\}.
\]
\end{lemma}
Finally, the IRWLS estimate of $\bpsi$ is derived by  repeatedly alternating
 the E-, S- and M-steps  until $| \ell(\bpsi^{(l+1)}) - \ell(\bpsi^{(l)})|$ becomes negligible. 

\subsection{Ridge Estimation Method}\label{sub:ridge}
Although the ML estimation is the common method to estimate the parameters of mixture of
 logistic regression models, the ML estimates are severely affected in presence of multicollinearity. 
 \cite{schaefer1984ridge} introduced  the ridge estimation as a proposal to rectify the multicollinearity problem. 
 One can obtain the ridge estimate ${\widehat\bpsi}_R$ by maximizing the ridge penalized log-likelihood 
 function of mixture of logistic regression models.
The ridge penalized log-likelihood function is given by
\begin{align}\label{loglik_ridge}
 \ell^R({\bf \beta}) = \ell({\bf \beta}) - \frac{1}{2} \lambda \bb^\top \bb,
 \end{align}
 where $\ell({\bf \beta})$ is the incomplete log-likelihood function \eqref{loglik_logistic}
  and $\lambda$ is the ridge parameter. Similar to Subsection \ref{sub:ls}, there is no closed 
  form for ${\widehat\bpsi}_R$ using \eqref{loglik_ridge}. We introduce the latent 
  variables ${\bf Z}=({\bf Z}_1,\ldots,{\bf Z}_M)$ and 
  develop again an SEM algorithm  on the complete data $({\bf X},{\bf y},{\bf Z})$ to obtain ${\widehat \bpsi}_{R}$.
  To do so, we implement the E-, S- steps of the ridge estimation method similar to Subsection \ref{sub:ls}. 
  
  In the M-step, the mixing proportion ${\widehat\pi}_j, j=1,\ldots,M$ is estimated from 
   \eqref{Jpihat_ls}. To estimate the coefficients, we maximize the 
   conditional expectation of the log-likelihood subject to the ridge penalty as
\begin{align}\label{JQ2_ridge}
{\bf Q}_2^R(\bb_j,\bpsi^{(l)}) = {\bf Q}_2(\bb_j,\bpsi^{(l)}) - \lambda_j \bb_j^\top \bb_j /2
\end{align}
where ${\bf Q}_2(\bb_j,\bpsi^{(l)})$ comes from \eqref{JQ2_ls} and $\lambda_j$ is the ridge 
parameter in $j$-th component of the mixture. 
\begin{lemma}\label{le:ridge}
Under the assumptions of Lemma \ref{le:ls}, the ridge estimator 
${\widehat\bb}_R^{(l+1)}=({\widehat\bb}_{R,1}^{(l+1)},\ldots,{\widehat\bb}_{R,M}^{(l+1)})$ using the IRWLS method is updated by
\[
{\widehat \bb}_{R,j} = \left({\bf X}_j^\top {\bf W}_j {\bf X}_j + \lambda_j \I \right)^{-1} 
{\bf X}_j^\top {\bf W}_j {\bf X}_j^\top {\widehat \bb}_{ML,j},
\]
where ${\widehat \bb}_{ML,j}$ is given by Lemma \ref{le:ls}.
\end{lemma}
Following \cite{inan2013liu}, we estimate the ridge parameter $\lambda_j$ by 
$\widehat\lambda_j = (p+1)/{\widehat \bb}_{ML,j}^\top {\widehat \bb}_{ML,j}$ where $p$ denotes the number of explanatory variables 
and ${\widehat \bb}_{ML,j}$ is the ML estimate of $\bb_j$. Finally, one can achieve  ${\widehat\bpsi}_R$ by  alternating
 the E-, S- and M-steps  until $| \ell(\bpsi_R^{(l+1)}) - \ell(\bpsi_R^{(l)})|$ becomes negligible. 
\subsection{Liu-type Estimation Method}\label{sub:liu}
When multicollinearity is severe, the ridge method may not be able to fully handle the sever 
ill-conditioned design matrix. 
\cite{liu2003using,inan2013liu} proposed the Liu-type (LT) method as a solution to the challenge 
in regression and logistics regression, respectively. 
We propose the LT  method in estimating  the parameters of the mixture of logistic 
regression models. To do that, we replace the ridge 
penalty $0= \lambda^{1/2} \bb + \epsilon'$ by the LT penalty 
\begin{align}\label{lt_penalty}
(-\frac{d}{\lambda^{1/2}}) {\widehat \bb} = \lambda^{1/2} \bb + \epsilon',
\end{align}
where ${\widehat \bb}$ can be any estimator of coefficients and $d \in \R$ and $\lambda > 0$ 
are two parameters of the LT estimation method. Throughout this manuscript, we use 
${\widehat \bb} = {\widehat \bb}_R$ in LT penalty \eqref{lt_penalty}.

In a similar vein to ML method (described in Subsection \ref{sub:ls}), we first view 
$({\bf X},{\bf y})$ as an incomplete data and translate them into complete data  $({\bf X},{\bf y},{\bf Z})$
 where ${\bf Z}$ include the missing component memberships.
We then use again SEM algorithm to find the LT estimate of the parameters of the 
mixture.  
Here, the E- and S-steps are handled like those of ML and ridge estimation methods.  

In the M-step, we first  use the classified data from S-step and estimate the mixing 
proportions from \eqref{Jpihat_ls}. Later, we maximize ${\bf Q}_2(\bb_j,\bpsi^{(l)})$ 
subject to LT penalty  \eqref{lt_penalty}  to estimate the coefficients within each 
partition ${P_j}^{(l+1)}$ for $j=1,\ldots,M$.
\begin{lemma}\label{le:lt}
Under the assumptions of Lemma \ref{le:ls}, the LT estimator 
${\widehat\bb}_{LT}^{(l+1)}=({\widehat\bb}_{LT,1}^{(l+1)},\ldots,{\widehat\bb}_{LT,M}^{(l+1)})$ 
using the IRWLS method is updated by
\[
{\widehat \bb}_{LT,j} = \left({\bf X}_j^\top {\bf W}_j {\bf X}_j + \lambda_j \I \right)^{-1} 
\left({\bf X}_j^\top {\bf W}_j {\bf V}_j + d_j {\widehat \bb}_{R,j}\right),
\]
where ${\bf W}_j$ and ${\bf V}_j$  are given by Lemma \ref{le:ls} and ${\widehat \bb}_{R,j}$ 
is calculated from Lemma \ref{le:ridge}.
\end{lemma}
Following  \cite{schaefer1984ridge}, there are various methods to estimate
$\lambda_j$. Here, we estimate the parameters by ${\widehat\lambda}_j= (p+1)/{\widehat\bb}_{R,j}^\top {\widehat\bb}_{R,j}$
 where $p$ is the number of explanatory variables and ${\widehat\bb}_{R,j}$ 
 denotes the ridge estimate of $\bb_j$.
 Once $\lambda_j$ are estimated, we use the operational technique of \cite{inan2013liu} and estimate the bias correction parameters $d_j$  by miximizing the mean square errors (MSE) of ${\widehat \bb}_{LT,j}$ within each partition $P_j^{(l+1)}$. It is easy to show that 
 \[
 \text{MSE}({\widehat \bb}_{LT,j}) = 
 tr\left[ \text{Var}({\widehat \bb}_{LT,j}) \right] +
 || \E({\widehat \bb}_{LT,j}) -\bb_j||_2^2, 
 \]
 where
 \begin{align*}
 tr \left[ \text{Var}({\widehat \bb}_{LT,j}) \right] = 
 tr &\left[  
 \left({\bf X}_j^\top {\bf W}_j {\bf X}_j + \lambda_j \I \right)^{-1}  
 \left({\bf X}_j^\top {\bf W}_j {\bf X}_j - d_j \I\right)
 \left({\bf X}_j^\top {\bf W}_j {\bf X}_j + \lambda_j \I \right)^{-1}  
 \left({\bf X}_j^\top {\bf W}_j {\bf X}_j \right) 
 \right. \\
 & \left. 
 \left({\bf X}_j^\top {\bf W}_j {\bf X}_j + \lambda_j \I \right)^{-1}
 \left({\bf X}_j^\top {\bf W}_j {\bf X}_j - d_j \I\right)
 \left({\bf X}_j^\top {\bf W}_j {\bf X}_j + \lambda_j \I \right)^{-1}
 \right],
 \end{align*}
 and 
 \begin{align*}
   ||\E({\widehat \bb}_{LT,j}) -\bb_j||_2^2  = 
 ||  
 \left({\bf X}_j^\top {\bf W}_j {\bf X}_j + \lambda_j \I \right)^{-1}  
 \left({\bf X}_j^\top {\bf W}_j {\bf X}_j - d_j \I\right)
 \left({\bf X}_j^\top {\bf W}_j {\bf X}_j + \lambda_j \I \right)^{-1}  
 {\bf X}_j^\top {\bf W}_j {\bf g}^{-1}({\bf X}_j;\bb_j) - \bb_j ||_2^2.
 \end{align*}
 As we can see, $ \text{MSE}({\widehat \bb}_{LT,j})$ depends on the true parameters $\bb_j$. Hence,
 the true  $\bb_j$ are replaced by ${\widehat \bb}_{R,j}$ in estimating the bias correction parameters of the LT method  
 ${\bf d} =(d_1,d_2,\ldots,d_M)$. 
 Finally, the E-, S- and M-steps of SEM algorithm under LT method is alternated
 until $| \ell(\bpsi_{LT}^{(l+1)}) - \ell(\bpsi_{LT}^{(l)})|$ becomes negligible.

\section{Simulation Studies}\label{sec:sim}
This section presents two simulation studies to compare the performance of the ML, Ridge and LT 
methods in estimating the parameters of the mixture of logistic regression models in the presence 
of multicollinearity. We investigate how the proposed estimation methods are affected by 
the sample size, multicollinearity level and the number of components in the mixture. 
We first consider that the underlying population is a mixture of two logistic regression 
models. The second simulation then studies the performance of the methods when the population 
comprises three logistic regression components.  

Following \cite{inan2013liu}, in the first simulation study, we used two parameters $\phi$ and $\rho$ to 
generate multicollinearity in the mixture of logistic regressions. We also considered the component logistic 
regressions include four covariates $({\bf x}_1,\ldots,{\bf x}_4)$ where $\phi^2$ and $\rho^2$ represent the 
association between the first and last two predictors in the mixture model. We first generated random 
numbers $\{w_{ij}, i=1,\ldots,n; j=1,\ldots,5\}$ from the standard normal distribution and then simulated the 
covariates by
\[
x_{i,j_1} = (1-\phi^2)^{1/2} w_{i,j_1} + \phi w_{i,5}, ~~~ j_1=1,2,
\]
\[
x_{i,j_2} = (1-\rho^2)^{1/2} w_{i,j_2} + \rho w_{i,5}, ~~~ j_2=3,4,
\]   
where we used $\phi = \{0.8,0.9,0.99\}$ and $\rho =\{ 0.9,0.99\}$ to simulate the multicollinearity in the mixture of logistic regressions.
 We then generated the binary responses form logistic regression 
$p_1({\bf x}_i;\bb_{01})$ with probability $\pi_0$ and from logistic regression $p_2({\bf x}_i;\bb_{02})$ 
with probability $1-\pi_0$ where $p_j(\cdot;\cdot)$ is given by \eqref{pj} and
$\Psi_0=(\pi_0,{\bb}_{01},{\bb}_{02})$ with $\pi_0=0.7$, ${\bb}_{01} = (1,3,4,5,6)$ and ${\bb}_{02} = (-1,-1,-2,-3,-5)$.

To examine the estimation performance of $({\widehat\pi},{\widehat\bb})$, we used the sum of squared errors (SSE) 
of the estimates and measured  $\sqrt{\text{SSE}({\widehat\bb})} = [({\widehat\bb}-{\bb}_{0})^\top ({\widehat\bb}-{\bb}_{0})]^{1/2}$ and 
$\sqrt{\text{SSE}({\widehat\pi})} = [({\widehat\pi}-{\widehat\pi}_{0})^\top ({\widehat\pi}-{\widehat\pi}_{0})]^{1/2}$
where ${\widehat\bb} = ({\widehat\bb}_1,{\widehat\bb}_2)^\top$ and ${\bb}_0 = ({\bb}_{01},{\bb}_{02})^\top$. 
To assess the classification performance of the methods,  we first estimated the parameters of the mixture 
model based on training sample of size $n$. We generated a validation set of size $100$ (independent from 
the training data) from the underlying mixture of two logistic regression. The trained model was then used 
to predict the binary response of the validation set.
We computed the prediction measures of Error = (FP+FN)/(TP+TN+FP+FN), 
Sensitivity = (TP)/(TP+FN) and Specificity = (TN)/(TN+FP) where FP, FN, TP and TN stand for false positive, false negative, 
true positive and true negative entries in the confusion matrix, respectively.  
We finally replicated 2000 times the estimation and prediction procedures using the ML, ridge and LT methods with sample size $n=\{25,100\}$. We computed 
the median and 95\% confidence interval (CI) of  $\sqrt{\text{SSE}}$, Error, Sensitivity and Specificity measures. We calculated the 
lower (L) and upper (U) bounds of the CI by 2.5 and 97.5 percentiles of the corresponding criterion, respectively.

\begin{table}[h]
\caption{\footnotesize{The median (M), lower (L) and upper (U) bounds of 95\% CIs for $\sqrt{\text{SSE}}$, Error, Sensitivity (SN) and Specificity (SP) of the methods 
in estimation and prediction of the mixture of two logistic regressions when $n=25$ and $\rho=0.9$.}}
\vspace{0.3cm} 
\centering 
\footnotesize{\begin{tabular}{ccccccccccccccccccc}
\hline 
 &  & \multicolumn{2}{c}{} & \multicolumn{3}{c}{$\sqrt{\text{SSE}}$} &  & \multicolumn{3}{c}{Error} &  & \multicolumn{3}{c}{SN} &  & \multicolumn{3}{c}{SP}\tabularnewline
\cline{5-7} \cline{9-11} \cline{13-15} \cline{17-19} 
$\phi$ & EM & $\ensuremath{\ensuremath{\boldsymbol{\Psi}}}$ &  & M & L & U &  & M  & L & U &  & M & L & U &  & M & L & U\tabularnewline
\hline 
0.85 & ML & $\ensuremath{\beta}$ &  & 223 & 46 & $1\times10^6$ &  & .46 & .28 & .68 &  & .55  & .20 & .83 &  & .55  & .19 & .83\tabularnewline
\cline{5-7} 
 &  & $\ensuremath{\pi}$ &  & .14 & .02 & .58 &  &  &  &  &  &  &  &  &  &  &  & \tabularnewline
\cline{2-19} 
 & Ridge & $\ensuremath{\beta}$ &  & 32  & 19 & 203 &  & .44  & .28 & .66 &  & .55  & .19 & .84 &  & .57  & .23 & .86\tabularnewline
\cline{5-7} 
 &  & $\ensuremath{\pi}$ &  & .22 & .02 & .70 &  &  &  &  &  &  &  &  &  &  &  & \tabularnewline
\cline{2-19} 
 & LT & $\ensuremath{\beta}$ &  & 30  & 21 & 36 &  & .46  & .30 & .60 &  & .56  & .29 & .81 &  & .55  & .26 & .78\tabularnewline
\cline{5-7} 
 &  & $\ensuremath{\pi}$ &  & .3  & .02 & .70 &  &  &  &  &  &  &  &  &  &  &  & \tabularnewline
\hline 
0.95 & ML & $\ensuremath{\beta}$ &  & 529  & 66 & $1\times10^6$ & & .46  & .28 & .68 &  & .55  & .21 & .83 &  & .56  & .21 & .84\tabularnewline
\cline{5-7} 
 &  & $\ensuremath{\pi}$ &  & .14  & .02 & .58 &  &  &  &  &  &  &  &  &  &  &  & \tabularnewline
\cline{2-19} 
 & Ridge & $\ensuremath{\beta}$ &  & 32  & 18 & 196 &  & .44  & .28 & .66 &  & .55  & .19 & .83 &  & .56  & .22 & .85\tabularnewline
\cline{5-7} 
 &  & $\ensuremath{\pi}$ &  & .22  & .02 & .70 &  &  &  &  &  &  &  &  &  &  &  & \tabularnewline
\cline{2-19} 
 & LT & $\ensuremath{\beta}$ &  & 30 & 21 & 36 &  & .46  & .30 & .60 &  & .55  & .30 & .81 &  & .54  & .27 & .79\tabularnewline
\cline{5-7} 
 &  & $\ensuremath{\pi}$ &  & .22  & .02 & .70 &  &  &  &  &  &  &  &  &  &  &  & \tabularnewline
\hline 
0.98 & ML & $\ensuremath{\beta}$ &  & 920 & 96 & $2\times10^6$ &  & .44 & .28 & .68 &  & .54  & .22 & .82 &  & .56 & .19 & .86\tabularnewline
\cline{3-7} 
 &  & $\ensuremath{\pi}$ &  & .14 & .02 & .58 &  &  &  &  &  &  &  &  &  &  &  & \tabularnewline
\cline{2-19} 
 & Ridge & $\ensuremath{\beta}$ &  & 31 & 19 & 207 &  & .44  & .26 & .68 &  & .56  & .19 & .83 &  & .56  & .23 & .85\tabularnewline
\cline{3-7} 
 &  & $\ensuremath{\pi}$ &  & .26 & .02 & .70 &  &  &  &  &  &  &  &  &  &  &  & \tabularnewline
\cline{2-19} 
 & LT & $\ensuremath{\beta}$ &  & 30 & 21 & 35 &  & .46  & .30 & .62 &  & .56  & .28 & .81 &  & .54 & .28 & .80\tabularnewline
\cline{3-7} 
 &  & $\ensuremath{\pi}$ &  & .24 & .02 & .70 &  &  &  &  &  &  &  &  &  &  &  & \tabularnewline
    \hline
    \end{tabular}}
         \label{2m_n25_r9}
\end{table}

Tables \ref{2m_n25_r9} and \ref{2m_n25_r99}-\ref{2m_n100_r99} show the results of the simulation study. 
We observe that the ML method performs marginally better than the ridge and LT methods in estimating 
the mixing proportions.
This relies on the fact that ridge and LT methods being considered biased shrinkage methods. By incorporating 
a bias into the estimation, these shrinkage methods are designed to overcome the multicollinearity 
and improve the analysis of the model's coefficients. While the ML estimates were dramatically 
affected by the multicollinearity, the ridge and LT estimates appeared significantly more reliable 
in estimating the coefficient of the mixture model. We also observe that ${\widehat\bb}_{LT}$ 
significantly outperforms ${\widehat\bb}_{R}$ where the CIs for $\sqrt{\text{SSE}}$ of ${\widehat\bb}_{R}$ 
account for 5-10 times wider than those of ${\widehat\bb}_{LT}$. Similar to the findings of \citep{inan2013liu},
  the classification performances of Error, Specificity and Sensitivity under the three methods are almost 
  the same. Interestingly, as the sample size decreases from 100 to 25, the LT shrinkage method appears more
   reliable than the ML and ridge in estimating the coefficients of all the logistic components. 

\begin{table}[h]
\caption{\footnotesize{The median (M), lower (L) and upper (U) bounds of 95\% CIs for $\sqrt{\text{SSE}}$, Error, Sensitivity (SN) and Specificity (SP) of the methods 
in estimation and prediction of the mixture of three logistic regressions when $n=50$.}}
\vspace{0.3cm} 
\centering 
\footnotesize{\begin{tabular}{ccccccccccccccccccc}
\hline 
 &  & \multicolumn{2}{c}{} & \multicolumn{3}{c}{$\sqrt{\text{SSE}}$} &  & \multicolumn{3}{c}{Error} &  & \multicolumn{3}{c}{SN} &  & \multicolumn{3}{c}{SP}\tabularnewline
\cline{5-7} \cline{9-11} \cline{13-15} \cline{17-19} 
$\phi$ & EM & $\ensuremath{\ensuremath{\boldsymbol{\Psi}}}$ &  & M & L & U &  & M  & L & U &  & M & L & U &  & M & L & U\tabularnewline
\hline 
0.85 & ML & $\ensuremath{\beta}$ &  & 139 & 32 & $6\times10^5$ &  & .45 & .32 & .65 &  & .60  & .18 & .84 &  & .48  & .20 & .77\tabularnewline
\cline{5-7} 
 &  & $\ensuremath{\pi}$ &  & .34 & .06 & .70 &  &  &  &  &  &  &  &  &  &  &  & \tabularnewline
\cline{2-19} 
 & Ridge & $\ensuremath{\beta}$ &  & 59  & 44 & 69 &  & .43  & .31 & .56 &  & .68  & .37 & .90 &  & .43  & .12 & .70\tabularnewline
\cline{5-7} 
 &  & $\ensuremath{\pi}$ &  & .44 & .14 & .80 &  &  &  &  &  &  &  &  &  &  &  & \tabularnewline
\cline{2-19} 
 & LT & $\ensuremath{\beta}$ &  & 60  & 46 & 65 &  & .42  & .30 & .55 &  & .69 & .38 & .93 &  & .42 & .12 & .72\tabularnewline
\cline{5-7} 
 &  & $\ensuremath{\pi}$ &  & .42 & .16 & .74 &  &  &  &  &  &  &  &  &  &  &  & \tabularnewline
\hline 
0.95 & ML & $\ensuremath{\beta}$ &  & 169 & 32 & $7\times10^5$ &  & .45 & .32 & .64 &  & .60 & .18 & .84 &  & .49 & .20 & .80\tabularnewline
\cline{5-7} 
 &  & $\ensuremath{\pi}$ &  & .32  & .06 & .70 &  &  &  &  &  &  &  &  &  &  &  & \tabularnewline
\cline{2-19} 
 & Ridge & $\ensuremath{\beta}$ &  & 58  & 43 & 68 &  & .43  & .31 & .55 &  & .67 & .38 & .91 &  & .44  & .15 & .72\tabularnewline
\cline{5-7} 
 &  & $\ensuremath{\pi}$ &  & .44 & .13 & .80 &  &  &  &  &  &  &  &  &  &  &  & \tabularnewline
\cline{2-19} 
 & LT & $\ensuremath{\beta}$ &  & 60 & 45 & 65 &  & .42  & .30 & .55 &  & .70  & .37 & .95 &  & .43  & .11 & .72\tabularnewline
\cline{5-7} 
 &  & $\ensuremath{\pi}$ &  & .42 & .16 & .72 &  &  &  &  &  &  &  &  &  &  &  & \tabularnewline
\hline 
0.99 & ML & $\ensuremath{\beta}$ &  & 255 & 38 & $2\times10^5$ &  & .45 & .32 & .64 &  & .61  & .20 & .84 &  & .49 & .22 & .79\tabularnewline
\cline{3-7} 
 &  & $\ensuremath{\pi}$ &  & .34 & .06 & .70 &  &  &  &  &  &  &  &  &  &  &  & \tabularnewline
\cline{2-19} 
 & Ridge & $\ensuremath{\beta}$ &  & 58 & 43 & 69 &  & .42  & .30 & .56 &  & .68  & .36 & .92 &  & .43  & .14 & .72\tabularnewline
\cline{3-7} 
 &  & $\ensuremath{\pi}$ &  & .42 & .14 & .67 &  &  &  &  &  &  &  &  &  &  &  & \tabularnewline
\cline{2-19} 
 & LT & $\ensuremath{\beta}$ &  & 60 & 45 & 65 &  & .42 & .29 & .55 &  & .70 & .34 & .95 &  & .43 & .09 & .76\tabularnewline
\cline{3-7} 
 &  & $\ensuremath{\pi}$ &  & .42 & .16 & .74 &  &  &  &  &  &  &  &  &  &  &  & \tabularnewline
    \hline
    \end{tabular}}
         \label{3m_n50}
\end{table}

The second simulation investigates the performance of the estimation methods when the population consists of 
three logistic regression models with two covariates. Assuming the correlation level $\phi=\{0.85,0.95,0.98\}$,  
we generated the covariates and binary responses as described above from the mixture population with 
${\bf\pi}_0=(0.3,0.4,0.3)$ and $\bb_{0}=(\bb_{01},\bb_{02},\bb_{03})$ with $\bb_{01}=(2.85,-10,-5.11)$, 
$\bb_{02}=(10,9.90,5.11)$ and 
 $\bb_{03}=(-3.84,9.90,5.11)$.
Similar to the setting of the first study, we computed the medians and 95\% CIs for the estimation and classification measures. 
The results of this study is presented in Tables \ref{3m_n50} and \ref{3m_n100}.
While the prediction performance of the three methods remains almost the same, the ridge and LT methods provided more reliable estimates 
for the coefficients of the mixture. 
Also, the LT estimates almost consistently outperform their ridge counterparts in estimating the coefficients.

\section{Bone Data Analysis}\label{sec:real}
Osteoporosis is a bone metabolic disease characterized by the deterioration of the bone tissues. 
Osteoporosis is typically called silent thief as it advances without any major symptoms. Approximately
 1 out of 3 women and 1 out of 5 men aged 50 and older experience osteoporosis \citep{melton1993long}.
  Bone mineral density (BMD) is considered as the gold standard in diagnosing the osteoporosis 
  \citep{world1994assessment,cummings1995risk}. Despite this reliability, BMD measurements are 
  obtained through an expensive and time-consuming process. BMD measurements, given as T-scores, 
  are obtained from dual X-ray absorptiometry images. When images are taken, medical experts are 
  needed to segment the images manually and find the final measurements. BMD scores are compared 
  with a BMD norm; the BMD mean of healthy individuals between 20 and 29 years old.
The bone status is diagnosed as osteoporosis when the BMD score is less than 2.5 standard deviation 
from the BMD norm.  

\begin{table}[h]
\caption{\footnotesize{The median (M), lower (L) and upper (U) bounds of 95\% CIs for $\sqrt{\text{SSE}}$, Error, Sensitivity (SN) and Specificity (SP) of the methods 
in the analysis of bone mineral data with sample size $n=\{20,40\}$.}}
\vspace{0.3cm} 
\centering 
\footnotesize{\begin{tabular}{ccccccccccccccccccc}
\hline 
 &  & \multicolumn{2}{c}{} & \multicolumn{3}{c}{$\sqrt{\text{SSE}}$} &  & \multicolumn{3}{c}{Error} &  & \multicolumn{3}{c}{SN} &  & \multicolumn{3}{c}{SP}\tabularnewline
\cline{5-7} \cline{9-11} \cline{13-15} \cline{17-19} 
$n$ & EM & $\ensuremath{\ensuremath{\boldsymbol{\Psi}}}$ &  & M & L & U &  & M  & L & U &  & M & L & U &  & M & L & U\tabularnewline
\hline 
20 & ML & $\ensuremath{\beta}$ &  & 7.8 & .70 & $6\times10^{51}$ &  & .36 & .20 & .64 &  & .00  & .00 & .81 &  & 1  & .19 & 1\tabularnewline
\cline{5-7} 
 &  & $\ensuremath{\pi}$ &  & .1 & .00 & .5 &  &  &  &  &  &  &  &  &  &  &  & \tabularnewline
\cline{2-19} 
 & Ridge & $\ensuremath{\beta}$ &  & 1.9 & .4 & 30.2 &  & .44  & .26 & .66 &  & .35  & .00 & .84 &  & .66  & .18 & 1\tabularnewline
\cline{5-7} 
 &  & $\ensuremath{\pi}$ &  & .2 & .00 & .7 &  &  &  &  &  &  &  &  &  &  &  & \tabularnewline
\cline{2-19} 
 & LT & $\ensuremath{\beta}$ &  & 1.9 & .58 & 3.2 &  & .46  & .28 & .62 &  & .33  & .00 & .78 &  & .64 & .28 & .97\tabularnewline
\cline{5-7} 
 &  & $\ensuremath{\pi}$ &  & .3  & .00 & .7 &  &  &  &  &  &  &  &  &  &  &  & \tabularnewline
\hline 
40 & ML & $\ensuremath{\beta}$ &  & 8.2 & 1.3 & $4\times10^{64}$ &  & .34 & .20 & .56 &  & .00 & .00 & .61 &  & 1 & .41 & 1\tabularnewline
\cline{5-7} 
 &  & $\ensuremath{\pi}$ &  & .07  & .00 & .45 &  &  &  &  &  &  &  &  &  &  &  & \tabularnewline
\cline{2-19} 
 & Ridge & $\ensuremath{\beta}$ &  & 1.8  & 1.2 & 26.4 &  & .44  & .28 & .62 &  & .33 & .00 & .75 &  & .67  & .33 & .97\tabularnewline
\cline{5-7} 
 &  & $\ensuremath{\pi}$ &  & .22 & .025 & .7 &  &  &  &  &  &  &  &  &  &  &  & \tabularnewline
\cline{2-19} 
 & LT & $\ensuremath{\beta}$ &  & 1.9 & 0.6 & 2.1 &  & .46  & .30 & .60 &  & .35  & .07 & .69 &  & .64  & .38 & .89\tabularnewline
\cline{5-7} 
 &  & $\ensuremath{\pi}$ &  & .3 & .05 & .7 &  &  &  &  &  &  &  &  &  &  &  & \tabularnewline
    \hline
    \end{tabular}}
         \label{bone_1}
\end{table}

Although measuring BMD scores is expensive, practitioners have access to various easily attainable 
characteristics about patients, such as physical and demographic characteristics and BMD results 
from earlier surveys.   Logistic regression is a practical statistical tool to use these characteristics 
to explain the osteoporosis stats of patients. The effect of these characteristics can vary within each 
osteoporosis class. Hence, one can use the mixture of logistic regressions to estimate the effects of these 
characteristics in an unsupervised learning approach.

This numerical study focused on the bone mineral data 
from the National Health and Nutritional Examination Survey (NHANES III). The Centers for Disease Control and 
Prevention (CDC) administered a survey to over 33999 American adults from 1988 to 1994. There are 182 
women aged 50 and older who participated in two bone examinations. Due to the significant impact 
of osteoporosis on the aged population of women, we treated these 182 women as our underlying population. 
 We used the total BMD from the second bone examination as our response variable and translated it into the 
 binary osteoporosis status. We also considered two easy-to-measure physical characteristics, including arm 
 and bottom circumferences, as two covariates of the logistic regressions. The high association between the 
 covariates $\rho=0.81$ indicates the multicollinearity problem in the mixture of logistic regressions. We 
 replicated 2000 times the ML, ridge and LT methods in estimating the parameters of the bone mineral population
 with training sample size $n=\{20,40,80,100\}$  and test sample size (taken independently from the training 
 step) of size $50$. We then computed the estimation and predication measures $\sqrt{\text{SSE}(\widehat\bb)}$,
  $\sqrt{\text{SSE}(\widehat\pi)}$, Error, Sensitivity, Specificity as described in Section \ref{sec:sim} where
   $\bb_0$ and $\pi_0$ are obtained by ML estimates of the parameters using the entire information of the population. 
 
 Tables \ref{bone_1} and \ref{bone_2} show the median (M) and 95\% confidence interval (CIs) 
 of the above estimation and prediction measures. The lower (L) and upper (U) bounds of the CIs were determined by 
 2.5 and 97.5 percentiles of the estimates. While the ML method slightly estimates  the mixing proportions better,  the ML method becomes extremely unreliable in estimating the coefficients of 
  component logistic regressions. Unlike the ML, the ridge and shrinkage methods could handle the multicollinearity issue in the estimation problem.  
 Comparing the shrinkages methods, ${\widehat\bb}_{LT}$ significantly outperforms ${\widehat\bb}_{R}$ in estimating the 
 coefficients of the mixture. 
Therefore,  the LT shrinkage method is recommended to estimate a mixture of logistic regressions when there is 
a multicollinearity in the analysis of bone mineral data.

\section{Summary and Concluding Remarks}\label{sec:sum}
In many medical applications, such as osteoporosis research, diagnosing the disease status requires
 an expensive and time-consuming process; however, practitioners have access to various easy-to-measure
  characteristics of patients, such as physical and demographic characteristics. Logistic regression is
   a powerful statistical method to take advantage of these characteristics to explain the disease status.
    When the population comprises several subpopulations, a mixture of logistic regressions enables us to
     investigate covariates' effect on the binary response in an unsupervised learning approach.

Although the Maximum likelihood (ML) method is the standard technique to estimate the parameter of a 
mixture of logistic regressions, 
the ML estimates are highly affected by multicollinearity.  
In this paper, we investigated the properties of the ridge and Liu-type (LT) shrinkage methods in estimating
 the mixture of logistic regressions. 
Through extensive numerical studies, we observed that the ML method slightly estimates better than shrinkage
 methods the mixing proportions of the mixture. As biased methods, shrinkage estimators are designed
  to overcome the ill-conditioned design matrix at the price of incorporating bias in the estimation. With
   multicollinearity, the ML method becomes extremely unreliable in estimating the coefficients of the mixture.
    Unlike the ML method, the proposed shrinkage methods provided reliable estimates. Comparing the shrinkage methods, 
${\widehat\bb}_{LT}$ outperforms considerably ${\widehat\bb}_{R}$ even in the presence of severe multicollinearity
 in the mixture of logistic regressions. Finally, we applied the proposed methods to bone mineral data to analyze
  the bone disorder status of women aged 50 and older.

\section*{Acknowledgment}  
Armin Hatefi and Hamid Usefi acknowledge the research support of the Natural Sciences and Engineering Research Council of Canada (NSERC).
\nocite{*}
\bibliographystyle{plainnat}
{\small\bibliography{main_sept4}
}

\section{Appendix}\label{app}

\subsection{ Proof of Lemma \ref{le:ls}}
From \eqref{grad_ls} and \eqref{hess_ls}, one can obtain ${\widehat \bb}^{(l+1)}_j$ based on partition $P_j^{(l)}$ by:
\begin{align*}
{\widehat\bb}_j^{(l+1)} &= 
{\widehat\bb}_j^{(l)} - {\bf H}_{\bb_j}^{-1}\left({\bf Q}_2(\bb_j,{\widehat\bpsi}^{(l)})\right) 
{\nabla_{\bb_j} {\bf Q}_2(\bb_j,{\widehat\bpsi}^{(l)})} \\
& = {\widehat\bb}_j^{(l)} +  \left({\bf X}_j^\top {\bf W}_j {\bf X}_j\right)^{-1}  {\bf X}_j^\top
\left[ {\bf y}_j - {\bf g}^{-1}({\bf X}_j;{\widehat\bb}_j^{(l)}) \right] \\
& =  \left({\bf X}_j^\top {\bf W}_j {\bf X}_j\right)^{-1}  {\bf X}_j^\top {\bf W}_j
\left\{{\bf X}_j {\widehat\bb}_j^{(l)} + {\bf W}^{-1}_j \left[ {\bf y}_j - {\bf g}^{-1}({\bf X}_j;{\widehat\bb}_j^{(l)}) \right] \right\}\\
& =   \left({\bf X}_j^\top {\bf W}_j {\bf X}_j\right)^{-1}  {\bf X}_j^\top {\bf W}_j {\bf V}_j.
\end{align*}
\hfill $\square$

\subsection{ Proof of Lemma \ref{le:ridge}}
Taking the first and second derivative from \eqref{JQ2_ridge} wrt $\bb_j$, the ridge gradient and ridge Hessian matrix are given by
\begin{align}\label{grad_ridge}
\nabla_{\bb_j} {\bf Q}^{R}_2(\bb_j,\bpsi^{(l)}) = {\bf X}_j^\top \left( {\bf y}_j - {\bf g}^{-1} ({\bf X}_j;\bb_j^{(l)}) \right) - \lambda_j \bb_j,
\end{align}
\begin{align}\label{hess_ridge}
{\bf H}_{\bb_j}\left({\bf Q}^{R}_2(\bb_j,\bpsi^{(l)})\right) = - {\bf X}_j^\top {\bf W}_j {\bf X}_j - \lambda_j \I.
\end{align}
Let ${\bf U}_j={\bf X}_j^\top {\bf W}_j {\bf X}_j + \lambda_j \I$. From \eqref{grad_ridge} and \eqref{hess_ridge}, the ridge estimate ${\widehat \bb}^{(l+1)}_{R,j}$ can be updated  by an iteratively re-weighted least squares as follows 
\begin{align*}
{\widehat\bb}_j^{(l+1)} &= 
{\widehat\bb}_j^{(l)} - {\bf H}_{\bb_j}^{-1}\left({\bf Q}^{R}_2(\bb_j,{\widehat\bpsi}^{(l)})\right) 
{\nabla_{\bb_j} {\bf Q}^{R}_2(\bb_j,{\widehat\bpsi}^{(l)})} \\
& = {\widehat\bb}_j^{(l)} + {\bf U}_j^{-1} 
\left\{  {\bf X}_j^\top
\left[ {\bf y}_j - {\bf g}^{-1}({\bf X}_j;{\widehat\bb}_j^{(l)}) \right]  - \lambda_j {\widehat\bb}_j^{(l)} \right\} \\
& = {\bf U}_j^{-1} {\bf U}_j {\widehat\bb}_j^{(l)} - \lambda_j {\bf U}_j^{-1} {\widehat\bb}_j^{(l)} +
{\bf U}_j^{-1}  {\bf X}_j^\top {\bf W}_j {\bf W}_j^{-1} \left[ {\bf y}_j - {\bf g}^{-1}({\bf X}_j;{\widehat\bb}_j^{(l)}) \right]\\
& = {\bf U}_j^{-1} {\bf X}_j^\top {\bf W}_j
\left\{{\bf X}_j  {\widehat\bb}_j^{(l)} + {\bf W}^{-1}_j \left[ {\bf y}_j - {\bf g}^{-1}({\bf X}_j;{\widehat\bb}_j^{(l)}) \right] \right\}\\
& = \left( {\bf X}_j^\top {\bf W}_j {\bf X}_j + \lambda_j \I\right)^{-1} {\bf X}_j^\top {\bf W}_j {\bf V}_j.
\end{align*}
\hfill $\square$

\subsection{ Proof of Lemma \ref{le:lt}}
It is easy to show that the gradient and ridge Hessian matrix under the LT estimation method are given by
\begin{align}\label{grad_lt}
\nabla_{\bb_j} {\bf Q}^{LT}_2(\bb_j,\bpsi^{(l)}) = {\bf X}_j^\top \left( {\bf y}_j - {\bf g}^{-1} ({\bf X}_j;\bb_j^{(l)}) \right) 
- d_j {\widehat\bb}_{R,j} - \lambda_j \bb_j,
\end{align}
\begin{align}\label{hess_lt}
{\bf H}_{\bb_j}\left({\bf Q}^{LT}_2(\bb_j,\bpsi^{(l)})\right) = - {\bf X}_j^\top {\bf W}_j {\bf X}_j - \lambda_j \I.
\end{align}
Let ${\bf U}_j={\bf X}_j^\top {\bf W}_j {\bf X}_j + \lambda_j \I$. From \eqref{grad_lt} and \eqref{hess_lt}, 
the LT estimate ${\widehat \bb}^{(l+1)}_{LT,j}$ can be updated  by an iteratively re-weighted least squares as follows 
\begin{align*}
{\widehat\bb}_j^{(l+1)} &= 
{\widehat\bb}_j^{(l)} - {\bf H}_{\bb_j}^{-1}\left({\bf Q}^{LT}_2(\bb_j,{\widehat\bpsi}^{(l)})\right) 
{\nabla_{\bb_j} {\bf Q}^{LT}_2(\bb_j,{\widehat\bpsi}^{(l)})} \\
&= {\widehat\bb}_j^{(l)} + {\bf U}_j^{-1} 
\left\{  {\bf X}_j^\top
\left[ {\bf y}_j - {\bf g}^{-1}({\bf X}_j;{\widehat\bb}_j^{(l)}) \right]  - \lambda_j {\widehat\bb}_j^{(l)}  
- d_j {\widehat\bb}_{R,j} \right\}\\
& = {\bf U}_j^{-1} {\bf U}_j {\widehat\bb}_j^{(l)} - \lambda_j {\bf U}_j^{-1} {\widehat\bb}_j^{(l)} +
{\bf U}_j^{-1}  {\bf X}_j^\top {\bf W}_j {\bf W}_j^{-1} \left[ {\bf y}_j - {\bf g}^{-1}({\bf X}_j;{\widehat\bb}_j^{(l)}) \right]
- d_j {\bf U}_j^{-1}  {\widehat\bb}_{R,j} \\
& = {\bf U}_j^{-1} {\bf X}_j^\top {\bf W}_j
\left\{{\bf X}_j  {\widehat\bb}_j^{(l)} + {\bf W}^{-1}_j \left[ {\bf y}_j - {\bf g}^{-1}({\bf X}_j;{\widehat\bb}_j^{(l)}) \right] \right\}
- d_j {\bf U}_j^{-1}  {\widehat\bb}_{R,j} \\
& = \left( {\bf X}_j^\top {\bf W}_j {\bf X}_j + \lambda_j \I\right)^{-1} 
\left\{ 
{\bf X}_j^\top {\bf W}_j {\bf V}_j - d_j {\widehat\bb}_{R,j} \right\}.
\end{align*}
\hfill $\square$


\newpage
\begin{table}[h]
\caption{\footnotesize{The median (M), lower (L) and upper (U) bounds of 95\% CIs for $\sqrt{\text{SSE}}$, Error, Sensitivity (SN) and Specificity (SP) of the methods 
in estimation and prediction of the mixture of two logistic regressions when $n=25$ and $\rho=0.99$.}}
\vspace{0.3cm} 
\centering 
\footnotesize{\begin{tabular}{ccccccccccccccccccc}
\hline 
 &  & \multicolumn{2}{c}{} & \multicolumn{3}{c}{$\sqrt{\text{SSE}}$} &  & \multicolumn{3}{c}{Error} &  & \multicolumn{3}{c}{SN} &  & \multicolumn{3}{c}{SP}\tabularnewline
\cline{5-7} \cline{9-11} \cline{13-15} \cline{17-19} 
$\phi$ & EM & $\ensuremath{\ensuremath{\boldsymbol{\Psi}}}$ &  & M & L & U &  & M  & L & U &  & M & L & U &  & M & L & U\tabularnewline
\hline 
0.85 & ML & $\ensuremath{\beta}$ &  & 1383 & 118 & $1\times10^6 $&  & .46 & .28 & .68 &  & .54  & .21 & .83 &  & .56  & .20 & .85\tabularnewline
\cline{5-7} 
 &  & $\ensuremath{\pi}$ &  & .14 & .02 & .58 &  &  &  &  &  &  &  &  &  &  &  & \tabularnewline
\cline{2-19} 
 & Ridge & $\ensuremath{\beta}$ &  & 32  & 19 & 181 &  & .44  & .28 & .68 &  & .56  & .19 & .84 &  & .56  & .23 & .86\tabularnewline
\cline{5-7} 
 &  & $\ensuremath{\pi}$ &  & .22 & .02 & .70 &  &  &  &  &  &  &  &  &  &  &  & \tabularnewline
\cline{2-19} 
 & LT & $\ensuremath{\beta}$ &  & 30  & 21 & 35 &  & .46  & .30 & .60 &  & .56  & .29 & .80 &  & .55  & .27 & .79\tabularnewline
\cline{5-7} 
 &  & $\ensuremath{\pi}$ &  & .3  & .02 & .70 &  &  &  &  &  &  &  &  &  &  &  & \tabularnewline
\hline 
0.95 & ML & $\ensuremath{\beta}$ &  & 1651  & 146 & $1\times10^6$ &  & .46  & .28 & .68 &  & .55  & .21 & .83 &  & .55  & .20 & .86\tabularnewline
\cline{5-7} 
 &  & $\ensuremath{\pi}$ &  & .14  & .02 & .54 &  &  &  &  &  &  &  &  &  &  &  & \tabularnewline
\cline{2-19} 
 & Ridge & $\ensuremath{\beta}$ &  & 31  & 19 & 171 &  & .44  & .28 & .68 &  & .56  & .21 & .84 &  & .56  & .21 & .86\tabularnewline
\cline{5-7} 
 &  & $\ensuremath{\pi}$ &  & .22  & .02 & .70 &  &  &  &  &  &  &  &  &  &  &  & \tabularnewline
\cline{2-19} 
 & LT & $\ensuremath{\beta}$ &  & 30 & 21 & 37 &  & .46  & .30 & .62 &  & .55  & .29 & .80 &  & .55  & .29 & .79\tabularnewline
\cline{5-7} 
 &  & $\ensuremath{\pi}$ &  & .3  & .02 & .70 &  &  &  &  &  &  &  &  &  &  &  & \tabularnewline
\hline 
0.98 & ML & $\ensuremath{\beta}$ &  & 2387 & 248 & $3\times10^6$ &  & .44 & .28 & .68 &  & .55  & .21 & .83 &  & .55 & .17 & .86\tabularnewline
\cline{3-7} 
 &  & $\ensuremath{\pi}$ &  & .14 & .02 & .58 &  &  &  &  &  &  &  &  &  &  &  & \tabularnewline
\cline{2-19} 
 & Ridge & $\ensuremath{\beta}$ &  & 31 & 18 & 326 &  & .44  & .26 & .68 &  & .56  & .19 & .83 &  & .57  & .22 & .85\tabularnewline
\cline{3-7} 
 &  & $\ensuremath{\pi}$ &  & .22 & .02 & .70 &  &  &  &  &  &  &  &  &  &  &  & \tabularnewline
\cline{2-19} 
 & LT & $\ensuremath{\beta}$ &  & 30 & 21 & 37 &  & .46  & .30 & .60 &  & .56  & .29 & .80 &  & .54 & .29 & .79\tabularnewline
\cline{3-7} 
 &  & $\ensuremath{\pi}$ &  & .3 & .02 & .70 &  &  &  &  &  &  &  &  &  &  &  & \tabularnewline
    \hline
    \end{tabular}}
         \label{2m_n25_r99}
\end{table}


\newpage
\begin{table}[h]
\caption{\footnotesize{The median (M), lower (L) and upper (U) bounds of 95\% CIs for $\sqrt{\text{SSE}}$, Error, Sensitivity (SN) and Specificity (SP) of the methods 
in estimation and prediction of the mixture of two logistic regressions when $n=100$ and $\rho=0.9$.}}
\vspace{0.3cm} 
\centering 
\footnotesize{\begin{tabular}{ccccccccccccccccccc}
\hline 
 &  & \multicolumn{2}{c}{} & \multicolumn{3}{c}{$\sqrt{\text{SSE}}$} &  & \multicolumn{3}{c}{Error} &  & \multicolumn{3}{c}{SN} &  & \multicolumn{3}{c}{SP}\tabularnewline
\cline{5-7} \cline{9-11} \cline{13-15} \cline{17-19} 
$\phi$ & EM & $\ensuremath{\ensuremath{\boldsymbol{\Psi}}}$ &  & M & L & U &  & M  & L & U &  & M & L & U &  & M & L & U\tabularnewline
\hline 
0.85 &ML & $\ensuremath{\beta}$ &  & 186 & 43 & $1\times10^4$ &  & .46 & .28 & .70 &  & .54  & .19 & .81 &  & .55  & .21 & .83\tabularnewline
\cline{5-7} 
 &  & $\ensuremath{\pi}$ &  & .12 & .01 & .65 &  &  &  &  &  &  &  &  &  &  &  & \tabularnewline
\cline{2-19} 
 & Ridge & $\ensuremath{\beta}$ &  & 30  & 20 & 52 &  & .44  & .30 & .60 &  & .57  & .33 & .80 &  & .56  & .30 & .78\tabularnewline
\cline{5-7} 
 &  & $\ensuremath{\pi}$ &  & .18 & .01 & .67 &  &  &  &  &  &  &  &  &  &  &  & \tabularnewline
\cline{2-19} 
 & LT & $\ensuremath{\beta}$ &  & 30  & 22 & 36 &  & .44  & .30 & .58 &  & .56  & .35 & .77 &  & .55  & .32 & .76\tabularnewline
\cline{5-7} 
 &  & $\ensuremath{\pi}$ &  & .26  & .01 & .70 &  &  &  &  &  &  &  &  &  &  &  & \tabularnewline
\hline 
0.95 & ML & $\ensuremath{\beta}$ &  & 291 & 66 & $2\times10^4$ &  & .46 & .28 & .68 &  & .54 & .19 & .81 &  & .56  & .21 & .85\tabularnewline
\cline{5-7} 
 &  & $\ensuremath{\pi}$ &  & .13  & .00 & .66 &  &  &  &  &  &  &  &  &  &  &  & \tabularnewline
\cline{2-19} 
 & Ridge & $\ensuremath{\beta}$ &  & 30  & 20 & 54 &  & .44  & .30 & .62 &  & .56 & .30 & .78 &  & .56  & .30 & .78\tabularnewline
\cline{5-7} 
 &  & $\ensuremath{\pi}$ &  & .21 & .01 & .68 &  &  &  &  &  &  &  &  &  &  &  & \tabularnewline
\cline{2-19} 
 & LT & $\ensuremath{\beta}$ &  & 29 & 22 & 35 &  & .44  & .30 & .60 &  & .57  & .33 & .78 &  & .56  & .33 & .76\tabularnewline
\cline{5-7} 
 &  & $\ensuremath{\pi}$ &  & .25 & .01 & .70 &  &  &  &  &  &  &  &  &  &  &  & \tabularnewline
\hline 
0.98 & ML & $\ensuremath{\beta}$ &  & 511 & 101 & $3\times10^5$ &  & .46 & .28 & .68 &  & .54  & .17 & .82 &  & .56 & .21 & .85\tabularnewline
\cline{3-7} 
 &  & $\ensuremath{\pi}$ &  & .12 & .00 & .65 &  &  &  &  &  &  &  &  &  &  &  & \tabularnewline
\cline{2-19} 
 & Ridge & $\ensuremath{\beta}$ &  & 30 & 19 & 52 &  & .44  & .28 & .61 &  & .57  & .32 & .79 &  & .56  & .30 & .79\tabularnewline
\cline{3-7} 
 &  & $\ensuremath{\pi}$ &  & .18 & .01 & .68 &  &  &  &  &  &  &  &  &  &  &  & \tabularnewline
\cline{2-19} 
 & LT & $\ensuremath{\beta}$ &  & 29 & 22 & 36 &  & .44  & .30 & .58 &  & .57 & .35 & .77 &  & .56 & .33 & .78\tabularnewline
\cline{3-7} 
 &  & $\ensuremath{\pi}$ &  & .24 & .01 & .70 &  &  &  &  &  &  &  &  &  &  &  & \tabularnewline
    \hline
    \end{tabular}}
         \label{2m_n100_r9}
\end{table}


\newpage
\begin{table}[h]
\caption{\footnotesize{The median (M), lower (L) and upper (U) bounds of 95\% CIs for $\sqrt{\text{SSE}}$, Error, Sensitivity (SN) and Specificity (SP) of the methods 
in estimation and prediction of the mixture of two logistic regressions when $n=100$ and $\rho=0.99$.}}
\vspace{0.3cm} 
\centering 
\footnotesize{\begin{tabular}{ccccccccccccccccccc}
\hline 
 &  & \multicolumn{2}{c}{} & \multicolumn{3}{c}{$\sqrt{\text{SSE}}$} &  & \multicolumn{3}{c}{Error} &  & \multicolumn{3}{c}{SN} &  & \multicolumn{3}{c}{SP}\tabularnewline
\cline{5-7} \cline{9-11} \cline{13-15} \cline{17-19} 
$\phi$ & EM & $\ensuremath{\ensuremath{\boldsymbol{\Psi}}}$ &  & M & L & U &  & M  & L & U &  & M & L & U &  & M & L & U\tabularnewline
\hline 
0.85 & ML & $\ensuremath{\beta}$ &  & 800 & 118 & $5\times10^4$ &  & .46 & .28 & .68 &  & .54  & .20 & .81 &  & .55  & .20 & .85\tabularnewline
\cline{5-7} 
 &  & $\ensuremath{\pi}$ &  & .12 & .00 & .64 &  &  &  &  &  &  &  &  &  &  &  & \tabularnewline
\cline{2-19} 
 & Ridge & $\ensuremath{\beta}$ &  & 30  & 19 & 59 &  & .44  & .28 & .62 &  & .56  & .30 & .78 &  & .56  & .30 & .80\tabularnewline
\cline{5-7} 
 &  & $\ensuremath{\pi}$ &  & .19 & .01 & .67 &  &  &  &  &  &  &  &  &  &  &  & \tabularnewline
\cline{2-19} 
 & LT & $\ensuremath{\beta}$ &  & 29 & 22 & 37 &  & .44  & .30 & .58 &  & .57  & .33 & .77 &  & .56 & .32 & .77\tabularnewline
\cline{5-7} 
 &  & $\ensuremath{\pi}$ &  & .25  & .01 & .68 &  &  &  &  &  &  &  &  &  &  &  & \tabularnewline
\hline 
0.95 & ML & $\ensuremath{\beta}$ &  & 951 & 176 & $6\times10^4$ &  & .46 & .28 & .70 &  & .54 & .19 & .81 &  & .56 & .19 & .85\tabularnewline
\cline{5-7} 
 &  & $\ensuremath{\pi}$ &  & .12  & .00 & .64 &  &  &  &  &  &  &  &  &  &  &  & \tabularnewline
\cline{2-19} 
 & Ridge & $\ensuremath{\beta}$ &  & 30  & 18 & 50 &  & .44  & .28 & .62 &  & .57 & .31 & .80 &  & .56  & .29 & .78\tabularnewline
\cline{5-7} 
 &  & $\ensuremath{\pi}$ &  & .23 & .01 & .68 &  &  &  &  &  &  &  &  &  &  &  & \tabularnewline
\cline{2-19} 
 & LT & $\ensuremath{\beta}$ &  & 29 & 21 & 36 &  & .44  & .30 & .58 &  & .57  & .35 & .78 &  & .56  & .33 & .76\tabularnewline
\cline{5-7} 
 &  & $\ensuremath{\pi}$ &  & .24 & .01 & .68 &  &  &  &  &  &  &  &  &  &  &  & \tabularnewline
\hline 
0.98 & ML & $\ensuremath{\beta}$ &  & 1331 & 268 & $8\times10^4$ &  & .46 & .28 & .70 &  & .54  & .21 & .81 &  & .56 & .20 & .84\tabularnewline
\cline{3-7} 
 &  & $\ensuremath{\pi}$ &  & .12 & .00 & .61 &  &  &  &  &  &  &  &  &  &  &  & \tabularnewline
\cline{2-19} 
 & Ridge & $\ensuremath{\beta}$ &  & 30 & 17 & 80 &  & .44  & .30 & .64 &  & .57  & .27 & .79 &  & .56  & .29 & .79\tabularnewline
\cline{3-7} 
 &  & $\ensuremath{\pi}$ &  & .20 & .01 & .68 &  &  &  &  &  &  &  &  &  &  &  & \tabularnewline
\cline{2-19} 
 & LT & $\ensuremath{\beta}$ &  & 29 & 20 & 37 &  & .44  & .30 & .58 &  & .57 & .33 & .77 &  & .56 & .33 & .77\tabularnewline
\cline{3-7} 
 &  & $\ensuremath{\pi}$ &  & .24 & .01 & .68 &  &  &  &  &  &  &  &  &  &  &  & \tabularnewline
    \hline
    \end{tabular}}
         \label{2m_n100_r99}
\end{table}


\newpage
\begin{table}[h]
\caption{\footnotesize{The median (M), lower (L) and upper (U) bounds of 95\% CIs for $\sqrt{\text{SSE}}$, Error, Sensitivity (SN) and Specificity (SP) of the methods 
in estimation and prediction of the mixture of three logistic regressions when $n=100$.}}
\vspace{0.3cm} 
\centering 
\footnotesize{\begin{tabular}{ccccccccccccccccccc}
\hline 
 &  & \multicolumn{2}{c}{} & \multicolumn{3}{c}{$\sqrt{\text{SSE}}$} &  & \multicolumn{3}{c}{Error} &  & \multicolumn{3}{c}{SN} &  & \multicolumn{3}{c}{SP}\tabularnewline
\cline{5-7} \cline{9-11} \cline{13-15} \cline{17-19} 
$\phi$ & EM & $\ensuremath{\ensuremath{\boldsymbol{\Psi}}}$ &  & M & L & U &  & M  & L & U &  & M & L & U &  & M & L & U\tabularnewline
\hline 
0.85 & ML & $\ensuremath{\beta}$ &  & 115 & 34 & $1\times10^4$ &  & .44 & .32 & .63 &  & .63 & .20 & .87 &  & .47  & .19 & .77\tabularnewline
\cline{5-7} 
 &  & $\ensuremath{\pi}$ &  & .38 & .07 & .82 &  &  &  &  &  &  &  &  &  &  &  & \tabularnewline
\cline{2-19} 
 & Ridge & $\ensuremath{\beta}$ &  & 58  & 44 & 68 &  & .42 & .31 & .54 &  & .68  & .42 & .91 &  & .43  & .15 & .70\tabularnewline
\cline{5-7} 
 &  & $\ensuremath{\pi}$ &  & .47 & .16 & .84 &  &  &  &  &  &  &  &  &  &  &  & \tabularnewline
\cline{2-19} 
 & LT & $\ensuremath{\beta}$ &  & 60  & 42 & 66 &  & .41  & .30 & .55 &  & .70 & .40 & .93 &  & .43 & .13 & .72\tabularnewline
\cline{5-7} 
 &  & $\ensuremath{\pi}$ &  & .45 & .18 & .79 &  &  &  &  &  &  &  &  &  &  &  & \tabularnewline
\hline 
0.95 & ML & $\ensuremath{\beta}$ &  & 131 & 37 & $1\times10^5$ &  & .44 & .31 & .63 &  & .64 & .20 & .87 &  & .47 & .18 & .76\tabularnewline
\cline{5-7} 
 &  & $\ensuremath{\pi}$ &  & .39  & .07 & .8 &  &  &  &  &  &  &  &  &  &  &  & \tabularnewline
\cline{2-19} 
 & Ridge & $\ensuremath{\beta}$ &  & 58  & 43 & 69 &  & .41  & .30 & .53 &  & .68 & .44 & .92 &  & .44  & .17 & .68\tabularnewline
\cline{5-7} 
 &  & $\ensuremath{\pi}$ &  & .48 & .16 & .84 &  &  &  &  &  &  &  &  &  &  &  & \tabularnewline
\cline{2-19} 
 & LT & $\ensuremath{\beta}$ &  & 60 & 43 & 66 &  & .41  & .29 & .53 &  & .71  & .42 & .94 &  & .43  & .13 & .71\tabularnewline
\cline{5-7} 
 &  & $\ensuremath{\pi}$ &  & .46 & .18 & .79 &  &  &  &  &  &  &  &  &  &  &  & \tabularnewline
\hline 
0.99 & ML & $\ensuremath{\beta}$ &  & 201 & 47 & $2\times10^4$ &  & .43 & .31 & .62 &  & .64  & .22 & .88 &  & .47 & .18 & .78\tabularnewline
\cline{3-7} 
 &  & $\ensuremath{\pi}$ &  & .38 & .08 & .81 &  &  &  &  &  &  &  &  &  &  &  & \tabularnewline
\cline{2-19} 
 & Ridge & $\ensuremath{\beta}$ &  & 58 & 43 & 68 &  & .41  & .30 & .54 &  & .68  & .43 & .92 &  & .45  & .16 & .69\tabularnewline
\cline{3-7} 
 &  & $\ensuremath{\pi}$ &  & .47 & .17 & .85 &  &  &  &  &  &  &  &  &  &  &  & \tabularnewline
\cline{2-19} 
 & LT & $\ensuremath{\beta}$ &  & 60 & 45 & 65 &  & .41 & .29 & .53 &  & .71 & .40 & .95 &  & .44 & .13 & .74\tabularnewline
\cline{3-7} 
 &  & $\ensuremath{\pi}$ &  & .47 & .18 & .79 &  &  &  &  &  &  &  &  &  &  &  & \tabularnewline
    \hline
    \end{tabular}}
         \label{3m_n100}
\end{table}


\newpage
\begin{table}[h]
\caption{\footnotesize{The median (M), lower (L) and upper (U) bounds of 95\% CIs for $\sqrt{\text{SSE}}$, Error, Sensitivity (SN) and Specificity (SP) of the methods 
in the analysis of bone mineral data with sample size $n=\{80,100\}$.}}
\vspace{0.3cm} 
\centering 
\footnotesize{\begin{tabular}{ccccccccccccccccccc}
\hline 
 &  & \multicolumn{2}{c}{} & \multicolumn{3}{c}{$\sqrt{\text{SSE}}$} &  & \multicolumn{3}{c}{Error} &  & \multicolumn{3}{c}{Sen} &  & \multicolumn{3}{c}{Spe}\tabularnewline
\cline{5-7} \cline{9-11} \cline{13-15} \cline{17-19} 
$n$ & EM & $\ensuremath{\ensuremath{\boldsymbol{\Psi}}}$ &  & M & L & U &  & M  & L & U &  & M & L & U &  & M & L & U\tabularnewline
\hline 
80 & ML & $\ensuremath{\beta}$ &  & 9.6 & 2.0 & $5\times10^{70}$ &  & .34 & .20 & .58 &  & .00  & .00 & .63 &  & 1 & .42 & 1\tabularnewline
\cline{3-7} 
 &  & $\ensuremath{\pi}$ &  & .05 & .00 & .42 &  &  &  &  &  &  &  &  &  &  &  & \tabularnewline
\cline{2-19} 
 & Ridge & $\ensuremath{\beta}$ &  & 1.8 & 1.2 & 8.9 &  & .44  & .30 & .60 &  & .33  & .00 & .67 &  & .68  & .45 & 1\tabularnewline
\cline{3-7} 
 &  & $\ensuremath{\pi}$ &  & .25 & .025 & .67 &  &  &  &  &  &  &  &  &  &  &  & \tabularnewline
\cline{2-19} 
 & LT & $\ensuremath{\beta}$ &  & 1.9 & 1.05 & 2.0 &  & .46  & .30 & .60 &  & .33 & .07 & .63 &  & .66 & .44 & .87\tabularnewline
\cline{3-7} 
 &  & $\ensuremath{\pi}$ &  & .27 & .025 & .7 &  &  &  &  &  &  &  &  &  &  &  & \tabularnewline
\hline 
100 & ML & $\ensuremath{\beta}$ &  & 8.9 & 2.0 & $2\times10^{73}$ &  & .34 & .20 & .56 &  & .00 & .00 & .64 &  & 1 & .41 & 1\tabularnewline
\cline{3-7} 
 &  & $\ensuremath{\pi}$ &  & .04 & .00 & .4 &  &  &  &  &  &  &  &  &  &  &  & \tabularnewline
\cline{2-19} 
 & Ridge & $\ensuremath{\beta}$ &  & 1.8 & .96 & 8.5 &  & .44 & .28 & .60 &  & .32 & .00 & .64 &  & .67 & .44 & .97\tabularnewline
\cline{3-7} 
 &  & $\ensuremath{\pi}$ &  & .26 & .02 & .67 &  &  &  &  &  &  &  &  &  &  &  & \tabularnewline
\cline{2-19} 
 & LT & $\ensuremath{\beta}$ &  & 1.9 & 1.1 & 2.0 &  & .46 & .30 & .60 &  & .33 & .08 & .64 &  & .65 & .45 & .85\tabularnewline
\cline{3-7} 
 &  & $\ensuremath{\pi}$ &  & .27 & .03 & .7 &  &  &  &  &  &  &  &  &  &  &  & \tabularnewline
    \hline
    \end{tabular}}
         \label{bone_2}
\end{table}

\end{document}